\documentclass
[aps,prl,twocolumn,showpacs,floats,amsmath]{revtex4}%
\usepackage{graphicx}
\usepackage{amsmath}
\usepackage{amsfonts}
\usepackage{amssymb}%
\usepackage{epsf}

\newcommand{\be}{\begin{equation}}
\newcommand{\ee}{  \end{equation}}
\newcommand{\ba}{\begin{eqnarray}}
\newcommand{\ea}{  \end{eqnarray}}

\begin{document}

\title{Gaussian Decoherence from Spin Environments} 
\author{W.H. Zurek}
\author{F.M. Cucchietti}
\author{J.P. Paz} 
\affiliation{Theoretical Division, MS B213, Los Alamos National 
Laboratory, Los Alamos, NM 87545} 
\date{February 10, 2004} 
 
\begin{abstract} 
We examine an exactly solvable model of decoherence -- a spin-system 
interacting with a collection of environment spins. We show that in
this model (introduced some time ago to illustrate environment--induced
superselection) generic assumptions about the coupling strengths lead to
a universal (Gaussian) suppression of coherence between pointer states.
We explore the regimes of validity of these results and discuss their relation to the
spectral features of the environment and to the Loschmidt echo (or fidelity).
Finally, we comment on the observation of such time dependence in spin echo 
experiments.
\end{abstract} 

\pacs{03.65.Yz;03.67.-a}

\maketitle 

A single spin--system ${\cal S}$ (with states $\left\{ \left|0\right>, \left|1\right> \right\}$)
interacting with an environment ${\cal E}$ of many independent spins
($\left\{ \left| \uparrow_k \right>, \left| \downarrow_k \right> \right\}$,
$k=1..N$) through
the Hamiltonian
\be
{\cal H_{SE}} =\left(\left|0\right>\left<0\right| - 
\left|1\right>\left<1\right| \right) \sum_{k=1}^N \frac{g_k}{2}
\left( \left| \uparrow_k \right> \left< \uparrow_k \right| - 
\left|\downarrow_k \right>\left<\downarrow_k \right| \right)
\label{hamiltonian0}
\ee
may be the simplest solvable model of decoherence in spin systems. 
It was introduced
some time ago \cite{Zurek82} to show that relatively 
straightforward
assumptions about the dynamics can lead to the emergence of a preferred
set of pointer states due to environment--induced superselection (einselection)
\cite{Zurek82,deco}.
Such models have gained additional importance in the past decade because of
their relevance to quantum information processing \cite{QIP}. 

The purpose of
our paper is to show that -- with a few additional natural and simple
assumptions --
one can evaluate the exact time dependence of the reduced density matrix, and
demonstrate that the off--diagonal components display a Gaussian (rather than
exponential) decay.
In effect, we exhibit a simple soluble example of a situation where the usual
Markovian \cite{Kossakowski} assumptions about the evolution of a quantum open
system are not satisfied at all times.
Apart from their implications for decoherence, our results are also relevant to quantum error
correction \cite{ErrorCorrection} where precise precise knowledge of the 
dynamics is essential to select an efficient strategy. 

To demonstrate the Gaussian time dependence of decoherence we first write down a
general solution for the model given by Eq.~(\ref{hamiltonian0}). 
Starting with:
\be
\left| \Psi_{\cal SE}(0)\right> = 
(a \left|0\right>+b \left|1\right>) \bigotimes_{k=1}^N
\left( \alpha_k \left| \uparrow_k \right> + 
\beta_k \left|\downarrow_k \right> \right),
\label{initialstate}
\ee
the state of ${\cal SE}$ at an arbitrary time is given by:
\be
\left| \Psi_{\cal SE}(t)\right> = 
a \left|0\right> \left|{\cal E}_0 (t)\right> 
+b \left|1\right>  \left|{\cal E}_1 (t)\right>
\label{phit}
\ee
where
\ba 
\left|{\cal E}_0 (t)\right> & = & \bigotimes_{k=1}^N
\left( \alpha_k e^{i g_k t/2} \left| \uparrow_k \right> 
+ \beta_k e^{-i g_k t/2} \left|\downarrow_k \right> \right)  \nonumber \\ 
&=& \left|{\cal E}_1 (-t)\right>.
\label{environ}
\ea
The reduced density matrix of the system is then:
\ba
\rho_{\cal S} & = & {\rm Tr} _{\cal E} \left| \Psi_{\cal SE}(t)\right> 
\left< \Psi_{\cal SE}(t)\right| \nonumber \\
& = & |a|^2 \left|0\right>\left<0\right|+ a b^{*} r(t)
\left|0\right>\left<1\right| \nonumber \\ 
& + & a^{*} b r^{*}(t) \left|1\right>\left<0\right| + |b|^2 \left| 1 \right> \left< 1 \right|,
\label{reducedrho}
\ea
where the {\it decoherence} factor
$r(t)=\left<{\cal E}_1 (t)|{\cal E}_0 (t)\right>$
can be readily obtained:
\ba
r(t)&=&\prod_{k=1}^N 
\left( |\alpha_k|^2 e^{i g_k t} + |\beta_k|^2 e^{-i g_k t} \right).
\label{roft}
\ea

It is straightforward to see that $r(0)=1$ and for $t>0$ it decays rapidly
to zero, so that the typical fluctuations of the off-diagonal terms of
$\rho_{\cal S}$ will be small for large environments, since:
\be
\left<|r(t)|^2 \right>=2^{-N} \prod_{k=1}^N \left( 1+(|\alpha_k|^2 - 
|\beta_k|^2)^2 \right),
\label{rsoft}
\ee
Here $\left<...\right>$ denotes a long time average \cite{Zurek82}.
Clearly, 
$\left<|r(t)|^2 \right> \underset{N\rightarrow \infty}{\longrightarrow} 0$,
leaving $\rho_{\cal S}$ approximately diagonal in a mixture of the pointer states 
$\left\{ \left|0\right>, \left|1\right> \right\}$ which retain preexisting
classical correlations.

This much was known since \cite{Zurek82}. The aim of this paper is to show
that, for a fairly generic set of assumptions, the form of $r(t)$ can be
further evaluated and that -- quite universally -- it turns out to be 
approximately Gaussian in time. Thus, the simple model of Ref. \onlinecite{Zurek82}
predicts a universal (Gaussian) form of the loss of quantum coherence, whenever
the couplings $g_k$ of Eq.~\ref{hamiltonian0} are sufficiently concentrated near
their average value so that their standard deviation
$\left<(g_k-\left<g_k\right>)^2\right>$ exists and is finite. When this condition
is not fulfilled other sorts of time dependence become possible. In particular,
$r(t)$ may be exponential when the distribution of couplings is a Lorentzian.

To obtain our main result we carry out the multiplication of Eq.(\ref{roft}), re--expressing
$r(t)$ as a sum:
\ba
r(t) &=& \prod_{k=1}^N |\alpha_k|^2  e^{i t \sum_{n} g_n} 
+ \sum_{l=1}^N |\beta_l|^2 \prod_{k\ne l}^N |\alpha_k|^2 \times \nonumber \\
& & e^{i t(-g_l +\sum_{n\ne l} g_n )} 
+ \sum_{l=1}^N\sum_{m\ne l}^N |\beta_l|^2 |\beta_m|^2 \times \nonumber \\
& &\prod_{k\ne l,m}^N |\alpha_k|^2 
e^{\left[i t (-g_l-g_m + \sum_{n\ne l,m}^N g_n )\right]} 
+...
\label{rexpansion}
\ea
There are $\binom{N}{0}$, $\binom{N}{1}$, $\binom{N}{2}$, ... etc. terms in the
consecutive sums above. 
The binomial pattern is clear, and can be made even more apparent by assuming 
that $\alpha_k=\alpha$ and $g_k=g$ for all $k$. Then,
\be
r(t)= \sum_{l=0}^N \binom{N}{l} |\alpha|^{2(N-l)} |\beta|^{2l} e^{ig(N-2l)t},
\label{binomial}
\ee
i.e., $r(t)$ is the binomial expansion of $r(t)=\left(|\alpha|^2 e^{igt}+
|\beta|^2 e^{-igt}\right)^N$. 

We now note that, as follows from the Laplace-de Moivre theorem \cite{Gnedenko}, 
for sufficiently large $N$ the coefficients of the binomial expansion
of Eq. (\ref{binomial}) can be approximated by a Gaussian:
\ba
\binom{N}{l} |\alpha|^{2(N-l)} |\beta|^{2l} \simeq 
\frac{\exp{\left[-\frac{(l-N|\beta|^2)^2}{2 N |\alpha \beta|^2}\right]}}
{\sqrt{2 \pi N |\alpha \beta|^2}} .
\label{gaussian}
\ea
This limiting form of the distribution of the eigenenergies 
of the composite ${\cal SE}$ system
immediately yields our main result: $r(t)$ is approximately Gaussian
since it is a Fourier transform of an approximately Gaussian distribution of 
the energies resulting from all the possible combinations of the couplings
with the environment. 

The set of all the resulting energies must have an 
(approximately) Gaussian distribution. This behavior is generic, a result of the
law of large numbers \cite{Gnedenko}:
these energies can be thought of as being the terminal points of an $N$--step random walk.
The contribution of the $k$--th spin of the environment to the random energy is $+g$ or
$-g$ with probability $|\alpha|^2$ or $|\beta|^2$ respectively (Fig. 1 a).

The same argument can be carried out in the more general case of Eq.~(\ref{rexpansion}).
The ``random walk'' picture that yielded the distribution of the couplings
remains valid (see Fig. 1 b).
However, now the individual steps in the random walk are not all equal. Rather,
they are given by the set $\left\{g_k \right\}$ (see Eq.~\ref{hamiltonian0}) 
with each step $g_k$ taken just once in a given walk.
There are $2^N$ such distinct random walks, each contributing with the weight given by 
the product of the relevant $|\alpha_k|^2$ and $|\beta_k|^2$ to 
the sum of Eq.~(\ref{rexpansion}). This exponential proliferation of the contributing
coupling energies allows one to anticipate rapid convergence to the universal Gaussian
form of the decoherence factor $r(t)$. 

\begin{figure}
\centering \leavevmode
\epsfxsize 3.2in
\epsfbox{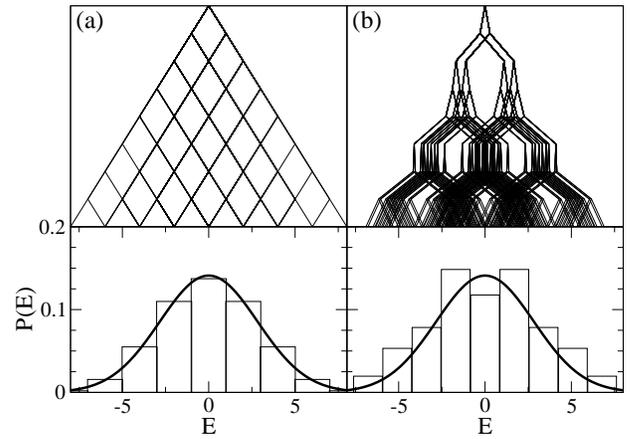}
\caption{The distribution of the energies obtains from the random walks with the
steps given by the coupling size and in the direction ($+g_k$ or $-g_k$) biased
by the probabilities $|\alpha_k|^2$ and $|\beta_k|^2$ as in Eq.~(\ref{Ldosrw})
(although in these examples we set $|\alpha_k|^2=1/2$).
(a) When all the couplings
have the same size $g_k=g$ (Eq.~(\ref{binomial})), 
a simple Newton's triangle leads to an approximate
Gaussian for the distribution of energies. (b) When the couplings
differ from step to step (Eq.~(\ref{rexpansion})), 
the resulting distribution still
has a approximately Gaussian envelope for large $N$. }
\label{Figure1} 
\end{figure}

Indeed, we can regard the energies resulting from the sums of $g_k$'s as a
random variable. Its probability distribution is given by products of the
corresponding weights. 
That is, the typical term in Eq.~\ref{rexpansion} is of the form:
\be
p_W e^{iE_W t} \equiv \left( \prod_{k\in W^+} |\alpha_k|^2e^{i g_kt}\right) 
 \left(\prod_{k\in W^-}|\beta_k|^2e^{-i g_kt} \right).
\ee
The resulting terminal energy is
\be
E_W=\sum_{k\in W^+} g_k-\sum_{k\in W^-} g_k,
\ee
and the cumulative weight $p_W$ is given by the corresponding product of
$|\alpha_k|^2$ and $|\beta_k|^2$. Each such specific random walk $W$
corresponding to a given combination of right ($k\in W^+$) and left ($k\in W^-$)
``steps" (see Fig. 1) contributes to the distribution of energies only once. 
The terminal points $E_{W}$ may or may not be degenerate: 
As seen in Fig. 1, in the degenerate case, the whole
collection of $2^N$ random walks ``collapses" into $N+1$ terminal
energies. More typically, in the degenerate case (also displayed in Fig. 1),
there are $2^N$ different terminal energies $E_W$. 
In both cases, the ``envelope" of the distribution $P(E_W)$ should be Gaussian, 
as we shall argue below.

We note
that the decoherence factor $r(t)$ can be viewed as the characteristic function
\cite{Gnedenko} (i.e., the Fourier transform)
of the distribution of eigenenergies $E_W$. Thus,
\be
r(t)=\int e^{iEt} \eta(E) dE,
\label{rLDOS}
\ee
where the strength function $\eta(E)$, also known as the local density of states (LDOS) 
\cite{LDOS} is defined in general as
\be
\eta(E)=\sum_\lambda |\left< \Psi_{\cal SE}(0)| \phi_\lambda \right>|^2 \delta(E-E_\lambda).
\ee
Above $\left|\phi_\lambda\right>$ are the eigenstates of the full Hamiltonian
and $E_\lambda$ its eigenenergies. In our particular model (Eq.~\ref{hamiltonian0}) 
the eigenstates are associated with all possible random walks in the set $W$, 
and therefore 
\be
\eta(E)=\sum_W p_W \delta(E-E_W).
\label{Ldosrw}
\ee
The discussion of decoherence in our model is thus directly related to 
the study of the characteristic function of the distribution of 
coupling energies $\eta(E)$.
Moreover, since the $E_W$'s are sums of $g_k$'s (that we assume independent of
each other), $r(t)$ is itself a product of characteristic functions of the distributions of
the couplings $\{g_k\}$, as we have already seen in the example of
Eq. (\ref{roft}). Thus, the distribution of $E_W$ belongs to the
class of the so--called {\it infinitely divisible distributions} \cite{Gnedenko,breiman}. 

\begin{figure}
\centering 
\epsfxsize 3.2in
\epsfbox{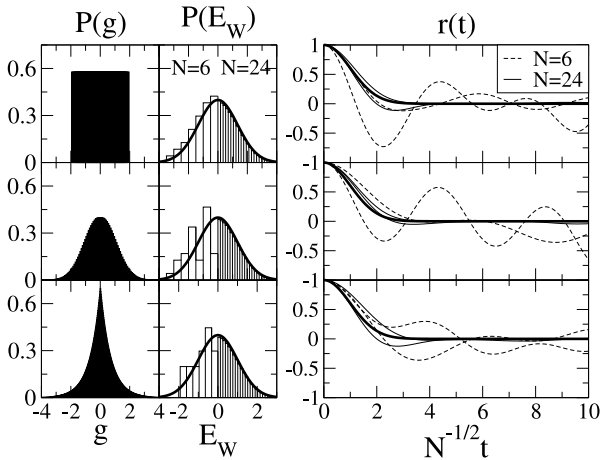}
\caption{(Left panels) Assumed distribution of the couplings $g_k$. 
(Center panels) Resulting distribution of the
eigenenergies $E_W$ (center panels) for $N=6$ ($E_W<0$) and $N=24$
($E_W>0$). In the case of $|\alpha_k|^2=1/2$ this distribution is in effect the
``strength function" (local density of states). (Right panels) 
Decoherence factor $r(t)$ for different initial
conditions with $N=6$ (dashed lines), $N=24$ (thin solid lines) and the average
(bold line).}
\label{Figure2} 
\end{figure}

The behavior of the decoherence factor $r(t)$ 
-- characteristic function of an infinitely divisible distribution --
depends only on the average and variance of the distributions of couplings
weighted by the initial state of the environment
\cite{Gnedenko,breiman}.
The remaining task is to calculate $\eta(E)$, which can be obtained through the
statistical analysis of the weighted random walk picture described above. If we
denote $x_k$ the random variable that takes the value $+g_k$ or $-g_k$ with probability
$|\alpha_k|^2$ or $|\beta_k|^2$ respectively, then its mean value
$a_k$ and its variance $b_k$ are
\ba
a_k&=&(|\alpha_k|^2-|\beta_k|^2)g_k, \nonumber \\
b_k^2&=&g_k^2-a_k^2=4|\alpha_k|^2|\beta_k|^2g_k^2.
\ea
The behavior of the sums of $N$ random variables $x_k$ (and thus, of their characteristic
function) depends on whether the so--called Lindeberg condition holds
\cite{Gnedenko}. It is expressed in terms of the cumulative variances
$B_N^2=\sum b_k^2$, and it is satisfied when the probability of the large
individual steps is small; e.g.:
\be
P(\underset{1\le k \le N}{{\rm max}} |g_k-a_k| \ge \tau B_N)
\underset{N\rightarrow\infty}{\longrightarrow} 0,
\ee
for any positive constant $\tau$. In effect, Lindeberg condition demands that $B_N$ be
finite: when it is met, the resulting distribution
of energies $E=\sum x_k$ is Gaussian
\be
P\left( \frac{E-{\overline E}_N}{B_N}<x\right) 
\underset{N\rightarrow\infty}{\longrightarrow}
\int_{-\infty}^{x} e^{-s^2/2}ds,
\ee
where ${\overline E}_N=\sum_k a_k$. In terms of the LDOS this implies
\be
\eta(E)\simeq \frac{1}{\sqrt{2\pi B_N^2}}
\exp{\left(\frac{-(E-{\overline E}_N)^2}{2 B_N^2}\right)},
\ee
an expression in excellent agreement with numerical results already for modest
values of $N$.
This distribution of energies yields a corresponding approximately
Gaussian time--dependence of $r(t)$, as seen in Fig. 2. Moreover, at least
for short times of interest for, say, quantum error correction, $r(t)$ is
approximately Gaussian already for relatively small values of $N$. This
conclussion holds whenever the initial distribution of the couplings has a
finite variance. The general form of $r(t)$ after applying the Fourier transform
of Eq. (\ref{rLDOS}) is
\be
r(t)\simeq e^{i {\overline E}_N t} e^{-B_N^2 t^2/2}.
\ee

It is also interesting to investigate cases when Lindeberg condition is not
met. Here, the possible limit distributions are given by the stable (or L\'{e}vy)
laws \cite{breiman}.
One interesting case
that yields an exponential decay of the decoherence factor corresponds to the
case of the Lorentzian distribution of couplings (see Fig. 3). 
Further intriguing questions concern the robustness of our conclusion under the
changes of the model. We shall address this issue elsewhere
\cite{CookPreparation} but, for the time being, we only note that the addition
of a strong self--Hamiltonian proportional to $\sigma_x$ changes the nature of the time
decay \cite{Dobrovitski}. On the other hand, small changes of the environment
Hamiltonians (like for instance dipolar interactions) 
seem to preserve the Gaussian nature of $r(t)$.

It is interesting to notice that the Fourier transform of the strength function
$\eta(E)$ is also related to the Loschmidt echo 
\cite{LETheo} (or fidelity) in the so called Fermi Golden
rule regime. The fact that the purity and the fidelity have closely related
decay rates has been recently shown \cite{LEdeco} 
for the case of a bath composed of non--interacting harmonic
oscillators. In this sense our results could be interpreted as an extension
of the discussion of Ref. \cite{LEdeco} to spin environments. 

The connection with fidelity is more easily seen if we write a
generalized version of the Hamiltonian (\ref{hamiltonian0}),
\ba
{\cal H_{SE}} =\frac{1}{2}\left(\left|0\right>\left<0\right| \otimes 
{\cal H}^0_{\cal E} + \left|1\right>\left<1\right| \otimes 
{\cal H}^1_{\cal E} \right) .
\label{hamiltonianCnot}
\ea
The decoherence factor is then the overlap of the initial state of the
environment $\left| \Psi_{\cal E}(0) \right>$ evolved with two different
Hamiltonians,
\ba
r(t)= \left< \Psi_{\cal E}(0) \right|e^{i{\cal H}^0_{\cal E} t/2} 
e^{-i{\cal H}^1_{\cal E} t/2} 
\left| \Psi_{\cal E}(0) \right>,
\ea
which clearly has the form of the amplitude of the Loschmidt echo for the
environment with the two states of the system as the perturbation. In the
particular model that we are treating, ${\cal H}_{\cal E}^0=-{\cal H}_{\cal E}^1$ and thus
\ba
r(t)&=& \left< \Psi_{\cal E}(0) \right|e^{-i{\cal H}_{\cal E}^1 t}
\left| \Psi_{\cal E}(0) \right>.
\label{autocorrelation}
\ea
This expression is the survival probability of the initial state of the
environment under the action of the Hamiltonian ${\cal H}_{\cal E}^1$, 
which has been shown to be the Fourier transform of the strength function
\cite{Heller}. This connection provides another way to understand
Eq.(\ref{rLDOS})

\begin{figure}
\centering \leavevmode
\epsfxsize 3.2in
\epsfbox{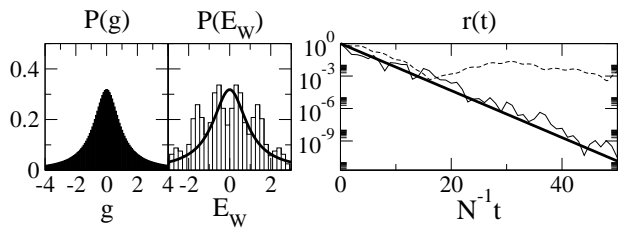}
\caption{Same as Fig. 2 but for a Lorentzian distribution of the couplings
${g_k}$. In this case $r(t)$ decays exponentially. 
The histogram and the dashed line in $r(t)$ correspond to $N=20$,
the straight thin line is a particular case for $N=100$ and the thick line is 
the average. We note that the convergence is slower than in the Gaussian case 
of Fig. 2, because realizations of ${g_k}$ are more likely to have 
one or two dominant couplings. Therefore, although the average shows a clear
exponential decay, fluctuations are noticeable even for large $N$.
Notice also that the logarithmic scale confirms the long time saturation of
$r(t)$ at $\sim 2^{-N/2}$, Eq~(\ref{rsoft}).}
\label{Figure3} 
\end{figure}

Possible experimental applications of our considerations are in nuclear
magnetic resonance, but also in other situations where two-level systems
interact with spin environments. We note that a Gaussian time dependence has
been seen in the NMR setting \cite{DiffusionNMR} but it is usually explained by
spin diffusion models (which have rather different character and employ a
different set of assumptions). 
Moreover, there is a substantial body of work \cite{Dobrovitski,deRaedt,Loss} 
on decoherence due to spin environments,
stimulated in part by the interests of quantum computation. 
The relation between the decoherence factor and the strength function might
prove useful in the physical setting of strongly 
interacting fermions, where it has been shown that the strength function takes a
Gaussian shape \cite{Kota}.
It is our hope that
the simple analytic model described here will assist in gaining further insights
into these fascinating problems.

We acknowledge fruitful discussions with R. Blume-Kohout and G. Raggio. We also
acknowledge partial support from ARDA/NSA grant. JPP received also partial support
from a grant by Fundaci\'on Antorchas.

\end{document}